\documentclass[prb,twocolumn,showpacs,amsmath,amssymb,superscriptaddress]{revtex4-1}
\usepackage[dvips]{graphicx}% Include figure files--uses eps figures
\usepackage{graphics}% Include figure file
\usepackage{dcolumn}% Align table columns on decimal point
\usepackage{bm}% bold math

\begin{document}

%\preprint{}

\title{Exchange bias of a ferromagnetic semiconductor by a ferromagnetic metal}


\author{K.~Olejnik}
\affiliation{Hitachi Cambridge Laboratory, Cambridge CB3 0HE, United Kingdom}
\affiliation{Institute of Physics ASCR, v.v.i., Cukrovarnicka 10, 16253 Praha 6, Czech Republic}
\author{P.~Wadley}
\author{J.~Haigh}
\author{K.~W.~Edmonds}
\author{R.~P.~Campion}
\author{A.~W.~Rushforth}
\author{B.~L.~Gallagher}
\author{C.~T.~Foxon}
\affiliation{School of Physics and Astronomy, University of Nottingham, Nottingham NG7 2RD, United Kingdom}
\author{T.~Jungwirth}
\affiliation{Institute of Physics ASCR, v.v.i., Cukrovarnicka 10, 16253 Praha 6, Czech Republic}
\affiliation{School of Physics and Astronomy, University of Nottingham, Nottingham NG7 2RD, United Kingdom}
\author{J.~Wunderlich}
\affiliation{Hitachi Cambridge Laboratory, Cambridge CB3 0HE, United Kingdom}
\affiliation{Institute of Physics ASCR, v.v.i., Cukrovarnicka 10, 16253 Praha 6, Czech Republic}
\author{S.~S.~Dhesi}
\author{S.~Cavill}
\author{G.~van~der~Laan}
\affiliation{Diamond Light Source, Harwell Science and Innovation Campus, Didcot, Oxfordshire, OX11 0DE, United Kingdom}
\author{E.~Arenholz}
\affiliation{Advanced Light Source, Lawrence Berkeley National Laboratory, Berkeley, California 94720, USA}

\date{\today}

\begin{abstract}
We demonstrate an exchange bias in (Ga,Mn)As induced by antiferromagnetic coupling to a thin overlayer of Fe. Bias fields of up to 240~Oe are observed. Using element-specific x-ray magnetic circular dichroism measurements, we distinguish a strongly exchange coupled (Ga,Mn)As interface layer in addition to the biassed bulk of the (Ga,Mn)As film. The interface layer remains polarized at room temperature.
\end{abstract}

%\keywords{}

\maketitle

Ferromagnetic (FM) semiconductors offer the prospect of combining high-density storage and gate-controlled logic in a single material. The realization of spin-valve devices from FM semiconductors requires the controlled switching of magnetization in adjacent layers between antiferromagnetic (AFM) and FM configurations. This has motivated several theoretical investigations of interlayer coupling in all-semiconductor devices \cite{AFMsuperlat}, and AFM coupling has recently been demonstrated in (Ga,Mn)As multilayers separated by $p$-type non-magnetic spacers \cite{Chung08}. However, the Curie temperature $T_C$ of (Ga,Mn)As is currently limited to 185~K in single layers \cite{Wang08}, and is typically much lower for layers embedded within a heterostructure \cite{Chung08}, which is an obstacle to the practical implementation of semiconductor spintronics.

The development of FM metal/FM semiconductor heterostructures has the potential to bring together the benefits of metal and semiconductor based spintronics, offering access to new functionalities and physical phenomena. Recent studies of MnAs/(Ga,Mn)As and NiFe/(Ga,Mn)As bilayer films have shown FM interlayer coupling and independent magnetization behavior, respectively \cite{Zhu07, Mark09}. Of particular interest is the Fe/(Ga,Mn)As system, since the growth of epitaxial Fe/GaAs(001) films is well-established \cite{Wastl05}. Remarkably, a recent x-ray magnetic circular dichroism (XMCD) study has shown that Fe may induce a proximity polarization in the near-surface region of (Ga,Mn)As, antiparallel to the Fe moment and persisting even above room temperature \cite{Maccher08}. Devices incorporating Fe/(Ga,Mn)As therefore offer the prospect of obtaining non-volatile room temperature spin-polarization in a semiconductor.

Until now, no information has been revealed about the coupling of Fe to (Ga,Mn)As layers away from the near-surface region. At the surface, the (Ga,Mn)As layer may be highly non-stoichiometric and Mn-rich, due to its non-equilibrium nature \cite{Campion03,Maccher06}. Previously, Fe/(Ga,Mn)As layers were produced by a process including exposure to air followed by sputtering and annealing prior to Fe deposition, which may further disrupt the interface order. The origin of the interface magnetism then had to be inferred by comparison to a series of reference samples \cite{Maccher08}. Demonstration of coupling between the bulk of the layers, {\it i.e.}, an exchange bias effect, would provide direct evidence of the interface magnetic order. Moreover, such coupling would offer new means of manipulating the FM semiconductor spin state and utilizing the proximity polarization effect in a spintronic device.

Here, we demonstrate an antiferromagnetic coupling and exchange bias in Fe/(Ga,Mn)As bilayer films, by combining element-specific XMCD measurements and bulk-sensitive superconducting quantum interference device (SQUID) magnetometry. As with previous studies of FM metal/FM semiconductor bilayers \cite{Zhu07, Mark09} (and in contrast to AFM coupled FM metal/FM metal exchange bias structures \cite{Binek06, Won07}) the layers are in direct contact without a non-magnetic spacer in between. We distinguish interface and bulk (Ga,Mn)As layers that are respectively strongly and weakly antiferromagnetically coupled to the Fe overlayer. In agreement with Ref.~\cite{Maccher08}, the interface layer remains polarized at room temperature.

The Fe and (Ga,Mn)As layers of the present study were both grown by molecular beam epitaxy in the same ultra-high vacuum system, in order to ensure a clean interface between them. The (Ga,Mn)As layer of thickness 10 to 50~nm was deposited on a GaAs(001) substrate at a temperature of 260$^\circ$C, using previously established methods \cite{Wang08, Campion03}. A low Mn concentration of $x\approx 0.03$ was chosen in order to avoid the formation of compensating Mn interstitials. The substrate temperature was then reduced to $\sim$0$^\circ$C, before depositing a 2~nm Fe layer, plus a 2~nm Al capping layer. In-situ reflection high energy electron diffraction and ex-situ x-ray reflectivity and diffraction measurements confirmed that the layers are single-crystalline with sub-nm interface roughness. SQUID magnetometry measurements were performed using a Quantum Design Magnetic Property Measurement System. Mn and Fe $L_{2,3}$ x-ray absorption and XMCD measurements were performed on beamline I06 at the Diamond Light Source, and on beamline 4.0.2 at the Advanced Light Source. Total-electron yield (TEY) and fluorescence yield (FY) were monitored simultaneously using the sample drain current and the photocurrent of a diode mounted at 90$^\circ$ to the incident beam, respectively.

SQUID magnetometry measurements were first performed on control Fe/GaAs(001) and (Ga,Mn)As/GaAs(001) samples, grown under the same conditions as the bilayers, to determine the magnetic anisotropies of the individual layers and the Curie temperature of the (Ga,Mn)As layer. The Fe film has a uniaxial magnetic anisotropy with easy axis along the [110] orientation, similar to previous studies \cite{Wastl05}. For the (Ga,Mn)As control sample, there is a competition between cubic and uniaxial magnetic anisotropies, with the former dominant at low temperatures and favoring easy axes along the in-plane $\langle 100 \rangle$ orientations, and the latter dominant close to $T_C$ ($\sim$35~K) giving an easy axis along the [1$\bar{1}$0] orientation. Figure~1 shows [110] magnetization versus temperature curves and low temperature hysteresis loops for a bilayer film containing a 20~nm thick (Ga,Mn)As layer. The total remnant moment of the bilayer film decreases on cooling under zero magnetic field below the $T_C$ of the (Ga,Mn)As, indicating that this layer aligns antiparallel to the Fe magnetization at zero field. The hysteresis curve shows a two-step magnetization reversal, indicating different behavior of the Fe and (Ga,Mn)As layers, with the smaller loop attributed to the dilute moment (Ga,Mn)As film. The minor hysteresis loop shown in Fig.~1 clearly shows a shift from zero field by a bias field $H_E$, indicating that the Fe layer induces an exchange bias in the magnetic semiconductor. The shape and size of the minor loop is in agreement with the hysteresis loop for the control (Ga,Mn)As sample, also shown in Fig.~1. This strongly indicates that the exchange bias affects the whole of the (Ga,Mn)As layer in the bilayer sample. 

Similar behavior is observed for bilayer samples containing a 10~nm or 50~nm (Ga,Mn)As layer, with a bias field which is approximately inversely proportional to the thickness $d$ of the ferromagnetic semiconductor layer (Fig.~1, inset). This 1/$d$ dependence of $H_E$ was found previously for MnAs/(Ga,Mn)As bilayers \cite{Zhu07}, and is generally observed in exchanged-biased thin films \cite{Nogues99}. From this dependence it is possible to describe the exchange bias in terms of an interface energy per unit area, $\Delta E=M_{FS}H_Ed=0.003$~erg/cm$^2$. This value is rather small compared to typical exchange bias systems \cite{Nogues99}, reflecting the low moment density $M_{FS}$ of the diluted FM semiconductor layer. However, the bias field for a given (Ga,Mn)As thickness is larger than is observed for MnO/(Ga,Mn)As structures \cite{Eid04}, while the reproducibility and flexibility of the present structures is much higher due to the single-crystalline ferromagnetic nature of the Fe layer.  

To confirm the presence of AFM interlayer coupling, we performed XMCD measurements at the Mn and Fe $L_{2,3}$ absorption edges in order to determine the magnetic response of the individual elements. In $L_{2,3}$ XMCD, electrons are excited from a 2$p$ core level to the unoccupied 3$d$ valence states of the element of interest by circularly polarized x-rays at the resonance energies of the transitions. The difference in absorption for opposite polarizations gives a direct and element-specific measurement of the projection of the 3$d$ magnetic moment along the x-ray polarization vector. The absorption cross-section is conventionally obtained by measuring the decay products -- either fluorescent x-rays or electrons -- of the photoexcited core hole. The type of decay product measured determines the probing depth of the technique. For Mn $L_{2,3}$ absorption, the probing depths for FY and TEY detection are $\lambda_{FY}\approx 100$~nm and $\lambda_{TEY}\approx 3$~nm. In the current experiment, the Mn XMCD measured using FY and TEY are thus sensitive to the bulk of the (Ga,Mn)As film and the near-interface layers, respectively. 

Figure~2(a)-(c) shows the magnetic field dependence of XMCD asymmetry, defined as $(I_l-I_r)/(I_l+I_r)$ where $I_{l(r)}$ is the absorption for left- (right-) circularly polarized x-rays. This is measured at the Fe and Mn $L_3$ absorption peaks for a Fe(2~nm)/(Ga,Mn)As(10~nm) sample at 2~K. The external field is applied along the photon incidence direction, which is at 70$^\circ$ to the surface normal with an in-plane projection along the [110] axis. The XMCD data show that the Fe film displays a square hysteresis loop with a single magnetization switch, as expected for a monocrystalline Fe film with strong uniaxial magnetic anisotropy. The Mn XMCD shows a more complicated loop due to the effect of the interlayer coupling. The projected Mn moment aligns antiparallel to the Fe moment at remanence, and undergoes a magnetization reversal of opposite sign to the Fe. With further increase of the external magnetic field, the Mn moment gradually rotates away from antiparallel alignment with the Fe layer, and into the field direction. Qualitatively similar behavior is observed for the Fe(2~nm)/(Ga,Mn)As(20~nm) sample: the (Ga,Mn)As layer is aligned antiparallel to the Fe layer at zero field, although the bias field is lower by approximately a factor of two.

Clear differences are observed between the Mn XMCD hysteresis loops obtained using TEY and FY detection modes. For FY the magnitude of the XMCD is similar (but of opposite sign) at remanence and at high magnetic fields, whereas for TEY at remanence it is approximately a factor of two larger than at 1000~Oe. The Mn $L_{2,3}$ XMCD spectra recorded at remanence and at 1000~Oe, shown in Fig.~3, confirm this result. At remanence the FY and TEY detected XMCD have similar magnitudes. However, under a large external field the XMCD is substantially smaller in TEY than in FY, confirming that the net magnetization of the Mn ions near the interface is significantly less than in the bulk of the (Ga,Mn)As film. This is the case even up to the highest field applied (20~kOe). By applying the XMCD sum rules \cite{sumrules} to the TEY data, and by comparing the spectra to previous measurements on well-characterized (Ga,Mn)As samples \cite{Jungw06}, the projected Mn 3$d$ magnetic moments are obtained as $-$1.4~$\mu_B$ and +0.8~$\mu_B$ per ion at remanence and 1000~Oe, respectively.

The difference between these values can be understood as being due to an interface layer which is strongly antiferromagnetically coupled to the Fe layer. At zero field, both the interfacial and bulk Mn are aligned antiparallel to the Fe layer. At high fields, the bulk of the (Ga,Mn)As layer away from the interface is re-oriented into the external field direction. However, the interfacial Mn remains antiparallel to the Fe layer and thus partially compensates the XMCD signal from the bulk of the (Ga,Mn)As. From the size of the remanent and 1000~Oe magnetic moments, it can be estimated that around 25-30\% of the TEY XMCD signal can be ascribed to the interfacial Mn which is strongly coupled to the Fe moments.

The interfacial Mn moments are ascribed to the proximity polarization of the (Ga,Mn)As interface by the Fe layer, such as was shown previously by XMCD as well as {\it ab initio} theory \cite{Maccher08}. Evidence for this can be observed from measurement of the Mn $L_{2,3}$ XMCD signal at temperatures above the (Ga,Mn)As $T_C$. Similar to the previous study \cite{Maccher08}, we observe a small but not negligible signal at room temperature (Fig.~3), with opposite sign to the Fe $L_{2,3}$ XMCD. Its spectral shape is characteristic of a localized electronic configuration close to $d^5$, similar to bulk (Ga,Mn)As \cite{Maccher08, Maccher06, Jungw06} but in contrast to Mn in more metallic environments such as Mn$_x$Fe$_{1-x}$ \cite{Maccher08} or MnAs \cite{Edmonds07}. A slight broadening is observed on the low energy side of the Mn $L_3$ peak, which may be due to the different screening induced by proximity to the Fe layer. Since the measured intensity is attenuated with distance $z$ from the surface as $I=I_0 \exp(-z/\lambda_{TEY})$, the thickness of the strongly coupled interface layer is estimated to be $\sim$0.7~nm or 2-3 monolayers, assuming a uniform distribution of Mn ions and magnetic moments throughout the (Ga,Mn)As film. This is around a factor of three thinner than in Ref.~\cite{Maccher08}, which could be due to the lower Mn concentration or the different preparation method of the present samples.

In summary, we have demonstrated antiferromagnetic coupling between Fe and (Ga,Mn)As layers in bilayer structures. A markedly different coupling is observed for the bulk of the (Ga,Mn)As layer and for Mn moments in the near-interface region. A thickness-dependent exchange bias field is observed to affect the whole of the bulk (Ga,Mn)As layer, which aligns antiparallel to the Fe layer at low fields, and switches to parallel when the external field is large enough to overcome the bias field and the magnetocrystalline anisotropy fields. In contrast, the interfacial Mn moments remain aligned antiparallel to the Fe layer even at 20~kOe, the largest field studied, and are polarized at temperatures well above the $T_C$ of the bulk (Ga,Mn)As layer. The latter observation confirms the recently reported result of Ref. 7, in which the Fe/(Ga,Mn)As bilayers were produced by a different method but showed qualitatively similar behavior of the interfacial moments. Our results shed new light on the magnetic coupling in Fe/(Ga,Mn)As hybrid layers which are of potential interest for room temperature spintronics, and also offer a means of controlling the spin orientation in a FM semiconductor.

% figures should be put into the text as floats.
% Use the graphics or graphicx packages (distributed with LaTeX2e)
% and the \includegraphics macro defined in those packages.
% See the LaTeX Graphics Companion by Michel Goosens, Sebastian Rahtz,
% entire page. There is no need to do explicit centering.

% Surround figure environment with turnpage environment for landscape
% figure
% \begin{figure}
% \end{figure}

%\begin{acknowledgments}
We acknowledge support from EU grants SemiSpinNet-215368 and NAMASTE-214499, and STFC studentship grant CMPC07100. The Advanced Light Source is supported by the U.S. Department of Energy under Contract No. DE-AC02-05CH11231. We thank Leigh Shelford for help during the Diamond beamtime.
%\end{acknowledgments}


\begin{figure}
\includegraphics[trim = 2mm 2mm 10mm 10mm,clip, width=80mm, angle=0]{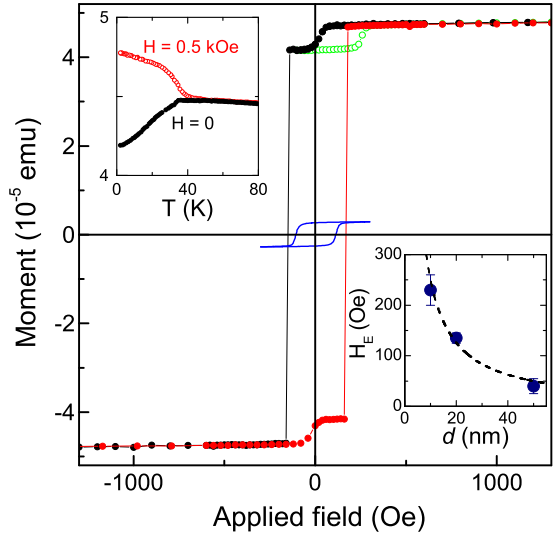}
\caption{\label{fig1}
(color) Main figure: Major (red/black) and minor (green) hysteresis loops along the [110] axis at 5~K, for a Fe (2~nm)/(Ga,Mn)As (20~nm) film, and the hysteresis loop for a control (Ga,Mn)As (20~nm) film along the same axis (blue). Left inset: Magnetization versus temperature for the Fe/(Ga,Mn)As film at remanence (black) and under a 500~Oe applied field (red). Right inset: Exchange bias field versus thickness $d$ of the (Ga,Mn)As film (points) and fit showing 1/$d$ dependence (dashed line).
}
\end{figure}

\begin{figure}
\includegraphics[trim = 2mm 2mm 10mm 10mm,clip, width=80mm, angle=0]{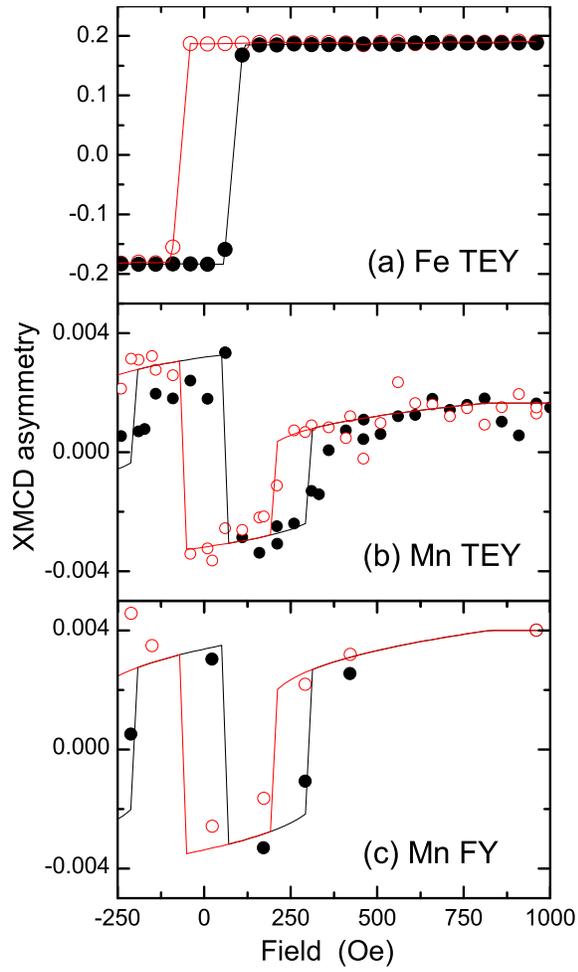}
\caption{\label{fig2}
(color online) XMCD asymmetry versus applied field along the [110] axis at 2~K, for a Fe (2~nm)/(Ga,Mn)As (10~nm) film. (a) Fe $L_3$, total electron yield; (b) Mn $L_3$, total electron yield; (c) Mn $L_3$, fluorescent yield. Black and red points are data for increasing and decreasing fields respectively; lines are to guide the eye.
}
\end{figure}

\begin{figure}
\includegraphics[trim = 2mm 2mm 10mm 10mm,clip, width=80mm, angle=0]{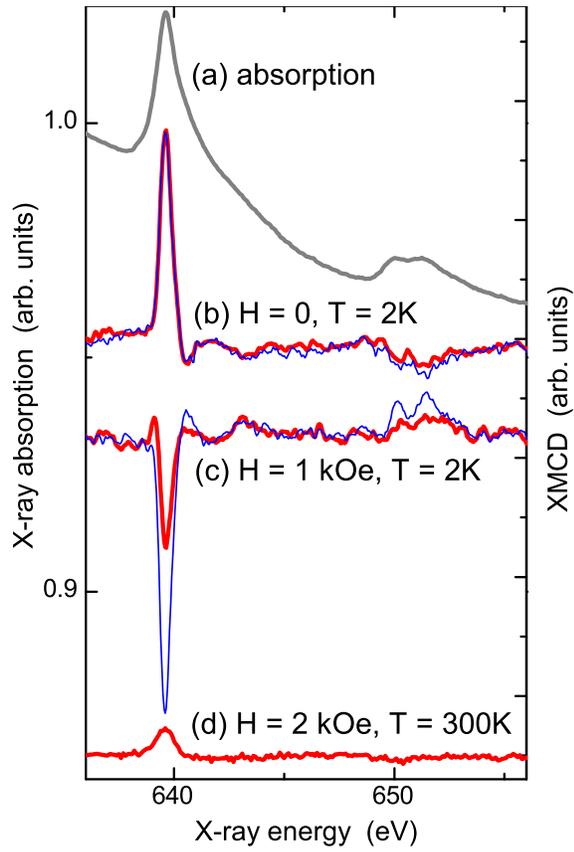}
\caption{\label{fig3} 
(color online) (a) Polarization-averaged Mn $L_{2,3}$ spectrum for a Fe/(Ga,Mn)As film; (b) XMCD spectra measured in remanence at 2~K; (c) XMCD spectra measured under a 1000~Oe applied field at 2~K; (d) XMCD spectrum measured under a 2000~Oe applied field at 300~K. XMCD spectra are obtained using TEY (thick red lines) and FY (thin blue lines) detection.
}
\end{figure}

\end{document}